\documentclass[aps,amsmath,amssymb,twocolumn,superscriptaddress]{revtex4}
\usepackage[]{graphicx}
\usepackage{color}
\usepackage{bbold}
\usepackage{braket}
\usepackage{multirow}
\usepackage[colorlinks=true, pdfstartview=FitV, linkcolor=blue, citecolor=red, urlcolor=black]{hyperref}

\newcommand{\be}{\begin{equation}}
\newcommand{\ee}{\end{equation}}
\newcommand{\ben}{\begin{eqnarray}}
\newcommand{\een}{\end{eqnarray}}
\newcommand{\bes}{\begin{subequations}}
\newcommand{\ees}{\end{subequations}}
\def\bal#1\eal{\begin{align}#1\end{align}}
\usepackage{comment}

\newcommand{\bfi}{\begin{figure}}
\newcommand{\efi}{\end{figure}}
\newcommand{\bc}{\begin{center}}
\newcommand{\ec}{\end{center}}

\begin{document}

\title{Kinks in generalized scalar field models and their scattering properties}
    \author{D. Bazeia}
    \affiliation{Departamento de F\'\i sica, Universidade Federal da Para\'\i ba, 58051-970 Jo\~ao Pessoa, PB, Brazil}
    \author{A. S. Lob\~ao}
    \affiliation{Escola T\'ecnica de Sa\'ude de Cajazeiras, Universidade Federal de Campina Grande, 58900-000 Cajazeiras, PB, Brazil}    
  \author{F. C. Simas}
    \affiliation{Programa de P\'os-Gradua\c c\~ao em F\'\i sica, Universidade Federal do Maranh\~ao\\Campus Universit\'ario do Bacanga, 65085-580, S\~ao Lu\'\i s, Maranh\~ao, Brazil}
    \affiliation{Departamento de F\'isica, Universidade Federal do Maranh\~ao (UFMA), Campus Universit\'ario do Bacanga, 65085-580, S\~ao Lu\'is, Maranh\~ao, Brazil}

\begin{abstract}
This work investigates kink solutions in one-dimensional scalar field theories. We begin with a review of the formalism used to obtain these solutions, presenting the BPS formalism and linear stability analysis. Next, we explore new models involving real scalar fields, generated by distinct potentials, with a focus on the topological structures responsible for the formation of kinks. Finally, we study collisions between the solutions obtained in two distinct models, analyzing their dynamic implications.
\end{abstract}

\maketitle
{I. \it Introduction. --}
Real scalar-field models are widely used as fundamental systems in the investigation of topological solutions in $(1+1)$ spacetime dimensions. Among these solutions, kinks stand out as classical, static configurations of the equations of motion, whose topology is determined by the field's boundary conditions, connecting different minima of the potential at the positive and negative spatial limits. The finite energy of these solutions requires that such minima correspond to distinct absolute values of the potential, which often implies the necessity of potentials with at least two distinct minima. This characteristic ensures the topological stability of kinks and allows their analysis in various physical contexts \cite{Manton:2004, Rajaraman:1982, Vilenkin:2000}.

For obtaining optimized solutions, the BPS formalism serves as an interesting strategy. By reformulating the equations of motion into a system of first-order differential equations, this technique ensures that BPS configurations minimize the system's energy, providing a robust foundation for the stability analysis of these solutions \cite{Bogomol, Prasad}. BPS solutions are highly significant, especially in systems exhibiting spontaneous symmetry breaking, bridging theoretical aspects with fields such as particle physics, cosmology and condensed matter physics \cite{Bazeia:2005tj,Bazeia:2007vx,Chiba:2014sda, Santos:2018hfg,Klimas:2019xnw,Adam:2018pvd}.

In this context, a common approach to deepen our understanding of kink dynamics is through the introduction of generalized models that enable new investigative features. For instance, we know that certain generalized models allow for the incorporation of diverse internal structures into physical quantities by simply modifying the shape of the potential \cite{Bazeia:2002xg,Bazeia:2003qt, Bazeia:2006pj,Buijnsters:2014,Chimento:2008ws,Bazeia:2017gub}. Additionally, these models can introduce unique and intricate behaviors into measured physical properties, further enriching the analysis and broadening the scope of potential applications. This approach has proven particularly promising, as modifying the potential can induce the emergence of new kink solutions that exhibit properties different from traditional kinks, such as compact kinks, half-compact, and long-range solutions \cite{Rosenau:1993zz,Dusuel:1998,Andrade:2024efr,Wobb}.

The study of generalized models by changing the potential has proven to be a highly productive field of investigation. Among the various possibilities, it has been shown for some time that the inclusion of the absolute value of the scalar field can significantly alter the behavior of kink solutions \cite{AAA,Bazeia:2001is}. In particular, the investigation of potentials involving the absolute value of the field has rekindled our attention. It has been shown that in a double-well potential, the absolute value creates a central hill, which impacts the scattering of kinks \cite{Karpisek:2024zdj}. This phenomenon has motivated us to explore generalized double-well potentials, as well as other potentials with multiple minima. Such configurations can exhibit characteristics that are quite distinct from traditional kinks, expanding the possibilities for analysis and opening new avenues for investigation.

Modifying the interaction potential not only affects the static properties of the kinks but can also alter the dynamics of their interactions. In this context, kink scattering, investigating the dynamic interactions of these solutions when subjected to perturbations or collisions, is a phenomenon that deserves detailed analysis. Numerical and analytical studies of kink scattering can reveal new behaviors, such as the formation of temporary bound states, radiation emission, and the emergence of fractal structures, which depend on the initial velocity and the nature of the involved potential. Phenomena such as resonance windows illustrate the exchange of energy between translational and vibrational modes, expanding the understanding of nonlinear interactions and providing a better understanding of the physics involved. 

Investigations on kink scattering of current interest appeared before in many works. Specifically, the first studies dedicated to examine kink-antikink inelastic collisions in the $\phi^4$ model were presented in Refs. \cite{Sugiyama:1979,Kudry,Getmanov}. The scattering process associated with this non-integrable model exhibits considerable complexity and richness, for instance, resonance windows can arise as a result of resonant energy exchange \cite{Campbell:1983}. Moreover, an important work \cite{Dorey:2011} identified a two-bounce window structure, even with the absence of a vibrational mode for an isolated kink or antikink. In Ref. \cite{Simas:2016} it was further shown that the appearance of extra shape modes causes a suppression of the resonant structure. Throughout this trend, multiple research studies have been conducted to investigate the interaction of kinks with other models \cite{Ivan:2019,Gani:2021,Azadeh,Casana:2024, Simas:2024}. In particular, we can mention wobbling kink scattering \cite{Alonso:2021,Azadeh:2021}, models with several scalar fields \cite{Halava:2011,Katsura} and scenarios in which the scalar field is in the quantum vacuum \cite{Mukho:2022}.  It is also worth noting the presence of spectral walls, which correspond to a barrier generated in the collision of kinks caused by the transition from discrete to continuous mode \cite{Adam:2019,Were}. Recently, Ref. \cite{Campos:2023} studied the kink-antikink collision in a model with an auxiliary function that alters the kinematics of one of the two scalar fields present in the model. Furthermore, there are investigations aimed at elucidating the kink collision process through the application of collective coordinates \cite{Takyi:2016,Oles:2021} and the study of oscillons \cite{Roman:2023,Blaschke:2024,Alexeeva:2024}, which appear naturally through the collapse of bubbles \cite{Gleiser:1995,khlopov}.

Motivated by the above investigations, in this paper we investigate kink solutions in one spatial dimension, with a particular focus on how modifications on the potential may influence the behavior of the kink solution, both in terms of its intrinsic characteristics and the scattering dynamics. The study is structured as follows: In Sec. II, we focus mainly on the methodology, reviewing the framework used to obtain kink solutions in one spatial dimension and presenting the BPS formalism. The investigation follows in Sec. III, exploring models that include the absolute value of the scalar field, generalizing the interaction potential. In Sec. IV, we study the kink scattering and discuss how the generalized models alter the kink-antikink collision output. Finally, in Sec. V, we provide some comments on the results and offer a brief discussion of future research directions.

{II. \it {Methodology. --}}
Kink-like solutions are localized structures that can be used as models to investigate various physical systems. In $1+1$ dimensions of spacetime, kink-like structures can be obtained from an action that describes a real scalar field in the form
\begin{equation}\label{action}
S=\int d^2x\,\left(\frac12\partial_\mu\phi\partial^\mu\phi-V(\phi)\right),
\end{equation}
where the potential $V(\phi)$ governs the self-interaction mechanism of the field $\phi$. The dynamics of the field is described by the equation of motion, which can be expressed as
\begin{equation}
\frac{\partial^2\phi}{\partial t^2}-\frac{\partial ^2\phi}{\partial x^2}+\frac{dV}{d\phi}=0.
\end{equation}
However, for static configurations, this equation reduces to
\begin{equation}\label{eqS}
\frac{d^2\phi}{dx^2}=V_\phi,
\end{equation}
where $V_\phi=dV/d\phi$. This differential equation is, in general, non-linear and depends on the potential $V(\phi)$. Therefore, obtaining interesting solutions often involves the careful choice or manipulation of the form of the potential. Nonetheless, if we want to investigate kinks with minimal energy, we can also rewrite the equation of motion in terms of the superpotential $W(\phi)$ as
\begin{equation}\label{eqF}
\frac{d\phi}{dx}= \pm W_\phi,
\end{equation}
where $V=(1/2)W_\phi^2$. This representation, known as the BPS formalism, allows the simplification of the equations of motion and the process of obtaining kink-like topological solutions by reformulating the equation of motion in terms of a superpotential $W(\phi)$. Instead of directly solving the nonlinear differential equation of the scalar field, the method reduces the problem to first-order equations, which are significantly easier to solve. This is achieved because the BPS formalism uses a decomposition that minimizes the energy of the solution, ensuring that the configurations obtained are minimum energy solutions, known as BPS solutions. Consequently, this formalism not only simplifies the mathematical analysis but also guarantees that the solutions have a clearer physical interpretation, tied to the stability and topological behavior of the system \cite{Bogomol,Prasad}. Here we can also refer to the result previously described in Ref. \cite{XXX}, which unveils the equivalence of Eq. \eqref{eqS} with Eqs. \eqref{eqF}.

In the BPS representation, we ensure topological behavior if, when $x \to \pm\infty$, the field goes as $\phi \to \bar{\phi}_{\pm}$, where $\bar{\phi}_{\pm}$ are vacuum values of $V(\phi)$ and $W(\bar{\phi}_+) \neq W(\bar{\phi}_-)$. The topological charge is given by
\begin{equation}
Q_T=W\,(\bar{\phi}_+)-W\,(\bar{\phi}_-).
\end{equation}
Additionally, we can express the energy density of the system in terms of the superpotential $W(\phi)$ as
\begin{equation}
\rho(x)=W_\phi^2.
\end{equation}
This causes the energy density to be directly determined by the behavior of the superpotential and the field solution at the vacuum values, given by $E = |Q_T|$. In addition, we can investigate the linear stability of the solution $\phi(x)$ by introducing small perturbations around it, in the form $\eta(x)\cos(\omega t)$. This procedure leads to a Schr\"odinger-like equation, given by
\begin{equation}
\left(-\frac{d^2}{dx^2}+U(x)\right)\eta(x)=\omega^2\eta(x),
\end{equation}
where
\begin{equation}\label{supepot}
U(x)=W_{\phi\phi}^2|_{\phi=\phi(x)}+\Big(W_{\phi}W_{\phi\phi\phi}\Big)|_{\phi=\phi(x)}.
\end{equation}
As we can see, the behavior of the model is strongly tied to the appropriate choice of the function $W(\phi)$, introduced to construct the BPS formalism. This approach is particularly useful when investigating generalized models in scalar field theory, where non-trivial potentials can give rise to a wide variety of solutions. By manipulating the form of the superpotential $W(\phi)$, one can explore different types of interactions and study how the specific choice of potential affects the behavior of the solutions. This opens up the possibility of constructing new models with rich and diverse solutions, allowing for a detailed investigation of phenomena such as the transition between compact and non-compact solutions, as well as the influence of perturbations and additional parameters. 

\begin{figure}[ht]
    \begin{center}
    \includegraphics[scale=0.9]{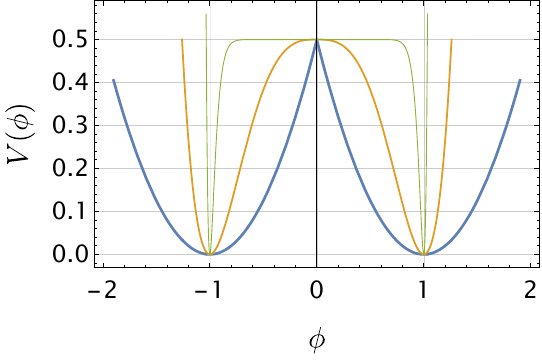}
    \end{center}
    \vspace{-0.5cm}
    \caption{\small{Potential $V(\phi)$ given by Eq. \eqref{Vmod1} for different values of $n$: $n=0$ (blue line), $n=1$ (orange line), and $n=10$ (light green line).}\label{fig1}}
\end{figure}

{III. {\it Models. --}} 
In recent years, several models have been developed to control the behavior of kink solutions. In particular, by modifying the superpotential, kink solutions have become increasingly diverse and intriguing. For example, we know that a superpotential in the form
\begin{equation}\label{mod1a}
W(\phi)=\phi-\frac{\phi^{2n+1}}{2n+1},
\end{equation}
where $n$ is an integer number, can transition between a compact and non-compact solution \cite{Bazeia:2014hja}. In a recent investigation \cite{Li:2024sqg}, it was also shown that the critical velocities in kink-antikink and antikink-kink collisions increased with $n$, with $W(\phi)$ in the  form
\begin{equation}\label{mod1b}
W(\phi)=\frac12\phi^2\left(1-\frac{\phi^{2n}}{n+1}\right).
\end{equation}
On the other hand, it was explored in \cite{Karpisek:2024zdj} that the presence of the absolute value of $\phi$ in the function $W$ has a strong impact on kink scattering by controlling the internal core structure of the solution. In \cite{Bazeia:2001is}, one of us demonstrated that the inclusion of the absolute value of the scalar field significantly alters the behavior of the system in field theories in $(1+1)$ dimensions. In that work, three models were investigated, two of which involved a single scalar field, while the other involved a combination of two interacting fields. In the single-field models, the superpotential was written in the form
\bes
\bal
W(\phi)&=\phi-\frac12\phi|\phi|,\label{mod2a}\\
W(\phi)&=\frac12\phi^2-\frac13\phi^2|\phi|.\label{mod2b}
\eal
\ees
The difference between them is that the first model has two minima at $\phi = \pm 1$ and a central peak at $\phi = 0$, whereas the second model has three minima: two at $\phi = \pm 1$ and one at $\phi = 0$. In this paper, we will follow the approach developed in \cite{Bazeia:2001is} and investigate how the introduction of the absolute value of the scalar field can modify the profile of the scalar field solution. To explore this, let us first consider the generalized model in the form
\begin{equation}\label{mod_A}
W(\phi)=\phi-\frac{\phi|\phi|^{2n+1}}{2n+2},
\end{equation}
where $n= 0, 1,2,\cdots$. The factor $2n\!+\!1$ was chosen to map only odd exponents; furthermore, if $n\!=\!0$, we recover the first model given by Eq. \eqref{mod2a}. With this we can obtain the potential $V(\phi)$ as
\begin{equation}\label{Vmod1}
V(\phi)=\frac12\left(1-|\phi|^{2n+1}\right)^2.
\end{equation}
This potential also features two minima separated by a maximum at $\phi=0$. However, when $n=0$, the maximum is represented by a sharp peak with a divergent first derivative. In contrast, for $n>0$, the maximum at $\phi=0$ becomes smoother, significantly altering the behavior of the kink solutions. This behavior is illustrated in Fig. \ref{fig1}, where we assign $n=0$ to the blue line, $n=1$ to the orange line, and $n=10$ to the light green line, representing the trend in the modification of the potential.

\begin{figure}[ht]
    \begin{center}
    \includegraphics[scale=0.9]{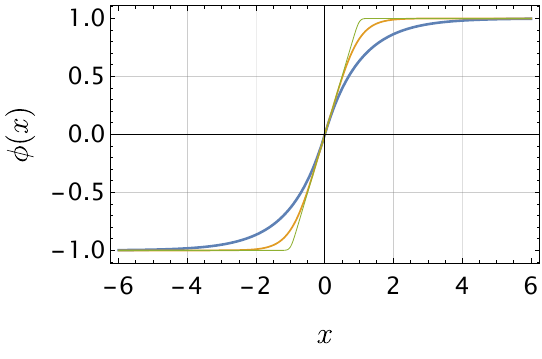}
    \end{center}
    \vspace{-0.5cm}
    \caption{\small{Kink solutions obtained numerically for $n=0$ (blue line), $n=1$ (orange line), and $n=10$ (light green line).}\label{fig2}}
\end{figure}

To obtain the solutions, we can employ the first-order formalism and express the first-order equation for the field $\phi$ as follows,
\begin{equation}
\frac{d\phi}{dx}=\pm\left(1-|\phi|^{2n+1}\right),
\end{equation}
where the plus/minus sign indicates the kink/antikink solution, respectively. Note that even with the use of the first-order formalism, the problem remains challenging, as the differential equation is difficult to solve for any value of $n$. However, we can express the solution as a transcendental equation that can be inverted using numerical methods. It is possible to verify that the solution has the form
\begin{equation}\label{solmod1}
\phi(x) \,\,{}_2F_1\Big(1;\frac1{2n+1};\frac{2n+2}{2n+1};|\phi(x)|^{2n+1}\Big)=\pm x,
\end{equation}
where ${}_2F_1$ denotes the hypergeometric function, defined by the power series as
\begin{equation}
{}_2F_1\Big(a;b;c;z\Big)=\sum_{n=0}^{\infty}\frac{a_nb_n}{c_n} \frac{z^n}{n!},
\end{equation}
with $|z|<1$. Fig. \ref{fig2} shows the kink solution obtained numerically for some values of $n$. As we can see, for increasing values of $n$ the kink solution becomes compact. As one knows, profiles with this characteristic were first highlighted in Ref. \cite{Rosenau:1993zz}. The compact solution is reduced to a straight line in the compact interval and behaves trivially outside of it. It is worth emphasizing that the compact kink energy density can only be obtained within the same closed interval, as opposed to the kink, whose density vanishes asymptotically.

As we can see, for increasing values of $n$ the kink solution becomes more and more compact, progressively narrowing to the region confined between $x = \pm 1$.

\begin{figure}[ht]
    \begin{center}
    \includegraphics[scale=0.9]{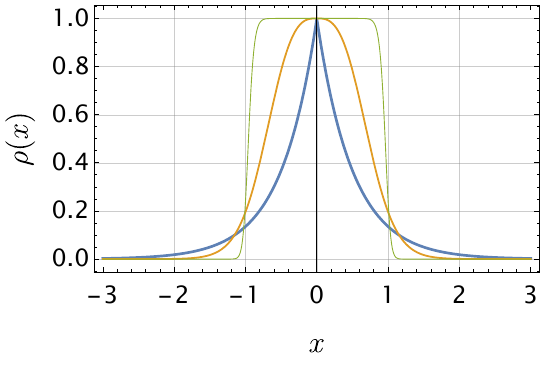}
    \end{center}
    \vspace{-0.5cm}
    \caption{\small{Energy density obtained for the model defined in Eq. \eqref{mod_A} for $n=0$ (blue line), $n=1$ (orange line), and $n=10$ (light green line).}\label{fig3}}
\end{figure}

To investigate the compactification process of the solutions, we also constructed the energy density shown in Fig. \ref{fig3}. As expected, the energy density exhibits a central peak at $x=0$ when $n=0$. As $n$ increases, this peak is smoothed out and the energy density tends to become concentrated in a compact region, resembling the behavior of a compact solution. We can show that for $n=0$, approximately $86.47\%$ of the energy is concentrated inside the region $x\in[-1,1]$. This percentage increases to $94.44\%$ when $n=1$, and further rises to $99.85\%$ for $n=10$. This trend indicates that as $n$ approaches infinity, the energy will become entirely concentrated within the compact region.

\begin{figure}[ht]
    \begin{center}
    \includegraphics[scale=0.9]{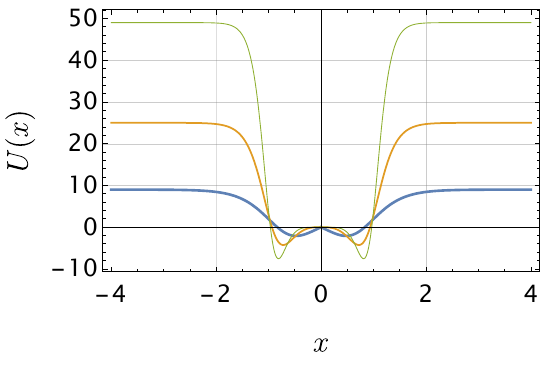}
    \end{center}
    \vspace{-0.5cm}
    \caption{\small{Stability potential for different values of $n$: $n=1$ (blue line), $n=2$ (orange line), and $n=3$ (light green line).}\label{fig4}}
\end{figure}

We can also use the superpotential defined in Eq. \eqref{mod_A} to obtain the stability potential \eqref{supepot}. However, because of the implicit form of the kink solution, which does not allow  for inversion in general, it is not possible to obtain the stability potential as a direct function of $x$ in this case. However, in the following we present the stability potential in terms of $\phi(x)$
\begin{equation}
U(x)=(2n+1)\Big((4n+1)\phi(x)^{4n}-2n|\phi(x)|^{2n-1}\Big),
\end{equation}
where $\phi(x)$ is given by Eq \eqref{solmod1}. It is important to note that the standard case with $n=0$, as investigated in \cite{Bazeia:2001is}, yields a $\delta(x)$ function. In this scenario, we have,
\begin{equation}
U(x)=1-2\delta(x).
\end{equation}
It was shown that the specific form of $U(x)$ for $n=0$ possesses only a single bound state, positioned at zero energy. On the other hand, for $n\geq 1$, we can construct confining potentials that do not involve the delta function. Fig. \ref{fig4} shows the stability potential for $n=1, 2$ and $n=3$. Interestingly, the perturbation potential for a single kink has a deep central well and its asymptotic values increase as $n$ increases. This implies in the presence of a greater number of bound states as $n$ increases to larger and larger values. 

In spite of the difficulty to describe the perturbation potential analytically, we have explored the occurrence of shape modes numerically.  We observed that the rise of $n$ is associated with the increase in internal modes. The numerical results are shown in Table I.


\begin{table}
\begin{tabular}{ p{3cm}|p{3cm}  }
\hline
\hline
\quad\quad\quad $n$ & \quad\quad\quad modes\\
\hline
\hline
\quad\quad\quad 1 & \quad\quad\quad\quad 2\\
\quad\quad\quad 2 & \quad\quad\quad\quad 5 \\
\quad\quad\quad 3 & \quad\quad\quad\quad 6 \\
\quad\quad\quad 4 & \quad\quad\quad\quad 9 \\
\quad\quad\quad 5 & \quad\quad\quad\quad 13 \\
\quad\quad\quad 6 & \quad\quad\quad\quad 17 \\ 
\hline
\hline
\end{tabular}
\caption{Number of vibrational modes for the kink obtained by model \eqref{mod_A}} 
\end{table}



\begin{figure}[ht]
    \begin{center}
    \includegraphics[scale=0.9]{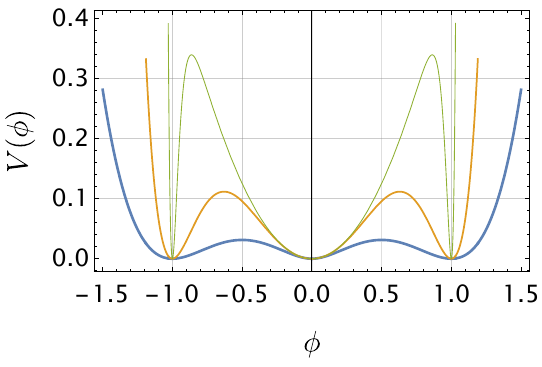}
    \end{center}
    \vspace{-0.5cm}
    \caption{\small{Potential $V(\phi)$ given by Eq. \eqref{Vmod_B} for different values of $n$: $n=0$ (blue line), $n=1$ (orange line), and $n=10$ (light green line).}\label{fig5}}
\end{figure}

Now, let us investigate how the inclusion of the absolute value can modify the behavior of a model with three minima. To do this, we will consider a modification of the model defined in \eqref{mod1b}, incorporating a coupling of the field $\phi$ with its absolute value in the following form,
\begin{equation}\label{mod_B}
W(\phi)=\phi^2\left(\frac12-\frac{|\phi|^{2n+1}}{2n+3}\right),
\end{equation}
where $n=0, 1, 2, \cdots$. It is interesting to note that in \eqref{mod1b}, the exponent results in an even value for any integer $n$. However, in \eqref{mod_B}, we desire the parameter $n$ to produce an odd exponent in the absolute value of the field. Furthermore, if $n=0$, we recover the second model given by Eq. \eqref{mod2b}. With the superpotential given by Eq. \eqref{mod_B}, we can derive the first-order equation as
\begin{equation}\label{FoEMB}
\frac{d\phi}{dx}=\pm\phi\left(1-|\phi|^{2n+1}\right).
\end{equation}
On the other hand, the interaction potential of the field can be expressed as follows,
\begin{equation}\label{Vmod_B}
V(\phi)=\frac12\phi^2\left(1-|\phi|^{2n+1}\right)^2.
\end{equation}
The graph presented in Fig. \ref{fig5} shows the potential $V(\phi)$ given by Eq. \eqref{Vmod_B}. As we can observe, the potential has a symmetric structure with respect to the origin, featuring multiple minima located at $\phi=\pm 1$ and at $\phi=0$. The behavior of the potential shares similarities with the $\phi^6$ model, particularly in how the curvature changes as the value of $\phi$ moves away from the minima. For small and large values of $n$, the behavior of the $\pm1$ minima of the potential is consistent with the behavior of compact solutions.

\begin{figure}[ht]
    \begin{center}
    \includegraphics[scale=0.9]{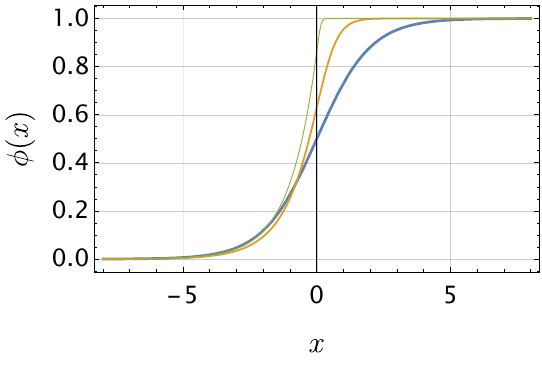}
    \end{center}
    \vspace{-0.5cm}
    \caption{\small{Kink solution given by Eq. \eqref{Solmod_B} for different values of $n$: $n=0$ (blue line), $n=1$ (orange line), and $n=10$ (light green line).}\label{fig6}}
\end{figure}

The solution obtained by integrating the first-order Eq. 
 \eqref{FoEMB} is
\begin{equation}\label{Solmod_B}
\phi(x)=\pm\left(\frac{e^{(2n+1)x}}{e^{(2n+1)x}+(2n+1)}\right)^{1/(2n+1)}.
\end{equation}
Note that we are only considering the kink/antikink solutions that connect the minima at $\phi=0$ and $\phi=\pm1$ of the potential. Fig. \ref{fig6} shows the kink solution obtained by Eq. \eqref{Solmod_B} for different values of $n$. We can observe that, for $n = 0$, represented by the blue line, the solution exhibits a smooth transition with no visible compactification. As the value of $n$ increases, indicated by the orange ($n = 1$) and light green ($n = 10$) lines, the transition becomes steeper, suggesting a partial compactification of the solution. This behavior resembles half-compact solutions, where the compactification occurs on only one side of the solution \cite{Bazeia:2020car, Bazeia:2023mat,Bazeia:2017gue,Bazeia:2015fca}. This is evidenced by the narrowing of the transition region as $n$ increases. For $n = 10$, for instance, the solution approaches a profile where the compactification manifests on one of the edges, while the other remains smooth, characterizing a kink solution that is compact on one side and extended on the other.

\begin{figure}[ht]
    \begin{center}
    \includegraphics[scale=0.9]{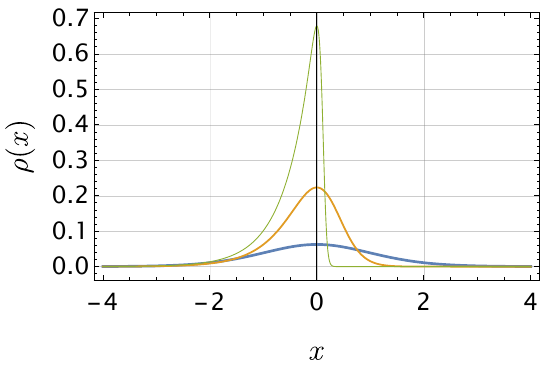}
    \end{center}
    \vspace{-0.5cm}
    \caption{\small{Energy density for the second model for different values of $n$: $n=0$ (blue line), $n=1$ (orange line), and $n=10$ (light green line).}\label{fig7}}
\end{figure}

Fig. \ref{fig7} shows the energy density corresponding to the second model for different values of $n$. Similarly to the compactification observed in the kink solution, the increase in $n$ intensifies the energy concentration around $x=0$, but notably, this compactification only occurs on one side, specifically for positive values of $x$. On the other side, for negative values of $x$, the energy density behaves in a more standard way, smoothly decaying without sharp peaks. For $n=0$, represented by the blue line, the energy density is well distributed across both sides. However, as $n$ increases, as seen in the orange line ($n=1$) and the light green line ($n=10$), the energy density becomes more compact and concentrated on the positive side, forming an increasingly sharp peak at $x=0$, while maintaining a smoother decay on the negative side. This asymmetric compactification appears to be induced by the absolute value of the field introduced in the investigated model.

For the second model, we can express the stability potential analytically, which is crucial for understanding the behavior of small perturbations around the kink solutions. The stability potential is derived from the second-order fluctuation equation and provides information into the spectrum of bound states and scattering modes that affect the overall stability of the kink. This analytical expression will be especially important in the next section, where we will investigate the scattering properties of the kink solutions. Understanding the stability potential allows us to predict how the kinks interact with external disturbances, which directly influences their dynamics in the scattering process. We can write that,
\begin{equation}
U(x)=1+a_n\phi(x)^{4n+2}-b_n|\phi(x)|^{2n+1},
\end{equation}
where $a_n=2(n+1)(4n+3)$ and $b_n=2(n+1)(2n+3)$. With this, substituting the solution $\phi(x)$ given by Eq. \eqref{Solmod_B}, we obtain,
\be
\begin{aligned}
U(x)=\,&1+a_n\frac{e^{2(2n+1)x}}{\left(e^{(2n+1)x}+(2n+1)\right)^{2}}\\
&-b_n\frac{e^{(2n+1)x}}{e^{(2n+1)x}+(2n+1)}.
\end{aligned}
\ee

\begin{figure}[ht]
    \begin{center}
    \includegraphics[scale=0.9]{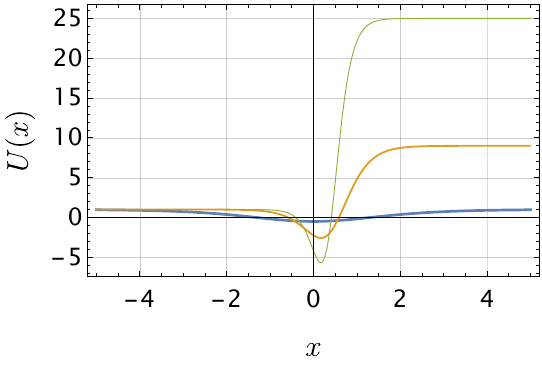}
    \end{center}
    \vspace{-0.5cm}
    \caption{\small{Schr\"odinger -like potential $U(x)$ for perturbations around the kink for the second model for different values of $n$: $n=0$ (blue line), $n=1$ (orange line), and $n=2$ (green line).}
    \label{fig8}}
\end{figure}

The perturbation potential for the kink for some values of $n$ is shown in Fig. \ref{fig8}. In this case, for $n=0$, there is a symmetric Schr\"odinger potential. However, increasing $n$ causes the appearance of an asymmetric potential. We can see that the asymptotic value for $U(x\rightarrow \infty)$ increases with $n$, while for $U(x\rightarrow -\infty)$, the value remains fixed. The potential for antikink can be found by changing $x \rightarrow -x$. 

The existence of bound states in the perturbation potential for the isolated kink or antikink has been investigated. For all values of the $n$ parameter, there is a zero mode, which corresponds to the translational mode. In the case of $n=0$, the potential favors the occurrence of the vibrational state. In fact, we obtain a bound state with frequency $\omega_1=0.866$, however, for $n\geq 1$, there are no shape modes. Due to the shape of the potential and the absence of bound states, the model discussed here is similar to the $\phi^6$ model \cite{Dorey:2011}. Therefore, we adopted the methodology of this study and examined the perturbations around the collective condition of the kink-antikink and antikink-kink pairs for $n= 1, 2, 3, \cdots$. As a result, the kink-antikink configuration has a central barrier that prevents the occurrence of bound states and hence the absence of two-bounce windows. Conversely, the Schr\"odinger-like potential for the antikink-kink composition has a deep central well, which permits the existence of a tower of bound states.


\begin{table}
\begin{tabular}{ p{3cm}|p{3cm}  }
\hline
\hline
\quad\quad\quad $n$ & \quad\quad\quad modes\\
\hline
\hline
\quad\quad\quad 1 & \quad\quad\quad\quad 5 \\
\quad\quad\quad 2 & \quad\quad\quad\quad 9 \\
\quad\quad\quad 3 & \quad\quad\quad\quad 12 \\
\quad\quad\quad 4 & \quad\quad\quad\quad 15 \\
\quad\quad\quad 5 & \quad\quad\quad\quad 18 \\
\quad\quad\quad 6 & \quad\quad\quad\quad 22 \\ 
\hline
\hline
\end{tabular}
\caption{Number of vibrational modes for the kink obtained by model \eqref{mod_B}. Here we have set $2x_0=16$ as the distance between the antikink and kink}.
\end{table}



We numerically solve the Schr\"odinger-like equation with the perturbation potential $U(x)$ to identify the bound states. Here, we examine pairs separated by a distance of $2x_0$. The potential minima, located at the antikink and kink positions, decrease as $n$ increases. Additionally, increasing $x_0$ leads to an increase in the number of vibrational modes. The Table II shows the main results for the occurrence of vibrational modes as a function of $n$, {\color{red} for $x_0 = 8$}. We noticed that the number of states increases significantly as the parameter value increases, which has a direct impact on the kink scattering process.

{IV. \it {Kink scattering. --}}
In this section, we will investigate kink scattering for models with two and three minima. We will discuss how the inclusion of the absolute value in the models and the variation of $n$ can alter the kink-antikink collision process.

{IV.A. {\it{ First model. --}}
Here we present the primary findings of the kink-antikink scattering process for the first model, which contains only two minima. For this, we solved the equation of motion using the five-point stencil with a spatial step $\delta x=0.05$ in a box $-z_{max}<x<z_{max}$, where we consider $z_{max}=200$. For resulting set of equations, we used the fifth-order Runge-Kutta method with an adaptive step size. Moreover, we fixed $x_0=\pm 10$ for the initial symmetric position of the pair and we have considered periodic boundary conditions. Therefore, for numerical solutions, we used the following initial conditions
\begin{eqnarray}
\phi(x,0,x_0,v) \!&=&\! \phi_{K}( x+ x_{0},0,v)  + \phi_{\bar{K}}( x-x_0,0,-v) -1 \text{,}\label{ic1} \nonumber
\\ 
\dot{\phi}(x,0,x_0,v) \!&=&\! \dot{\phi}_{K}( x+ x_{0},0,v ) + \dot{\phi}_{\bar{K}}(x-x_0,0,-v )\text{,} \nonumber
\end{eqnarray}
where $\phi_K(x,t,v) = \phi_{K}\big(\gamma(x-vt) \big)$ means a boost of Lorentz for the static kink solution, with $\gamma=(1-v^2)^{-1/2}$ and $v$ representing the velocity. Our numerical results are presented in the discussion that follows. We investigate collisions varying the initial velocity $v$ and the parameter $n$. First, we examined the outcomes for $n=0$. We examined various velocities values and chose $v=0.20$ as a representative case, whose kink-antikink scattering is shown in Fig. \ref{colA}. We observed that the kink-antikink pair is annihilated after the interaction. This result is related to the fact that the model only has the translational mode when $n=0$.

\begin{figure}[ht]
    \begin{center}
    \includegraphics[scale=0.46]{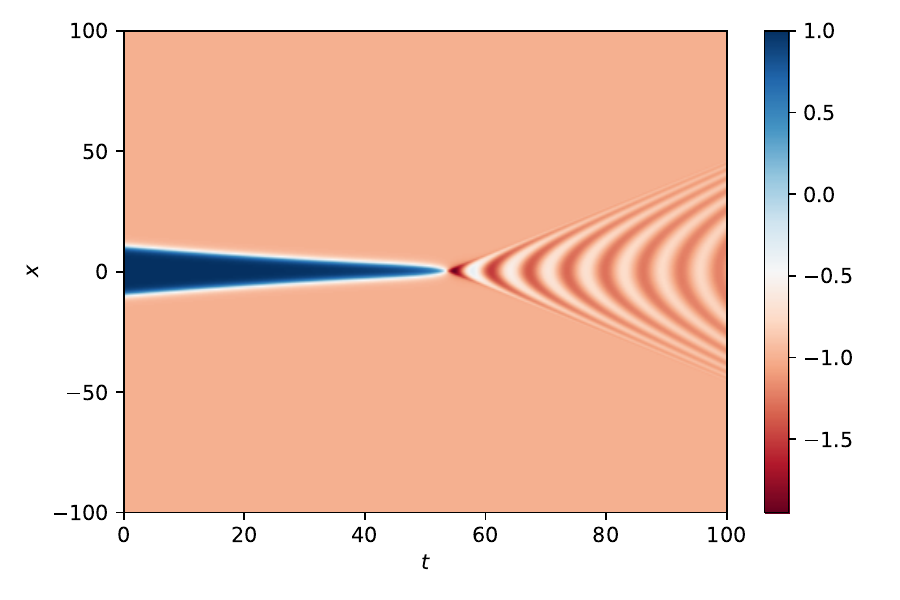}
    \end{center}
    \vspace{-0.5cm}
    \caption{\small{Kink-antikink scattering: field evolution in spacetime for $n=0$ and $v=0.20$.}
    \label{colA}}
\end{figure}

Furthermore, we investigated scattering for increasing values of the parameter $n$. For example, we can see in Fig. \ref{colB} the outcome of a collision for $n=1$ for different velocities. The bion and one-bounce behavior are displayed in Figs. \ref{colB}a and \ref{colB}b, respectively. This $n$ region corresponds to the perturbation potential transition, when the presence of bound states is detected, influencing the collision outcome via the resonant energy exchange process.

\begin{figure}[ht]
    \begin{center}
    \includegraphics[scale=0.28]{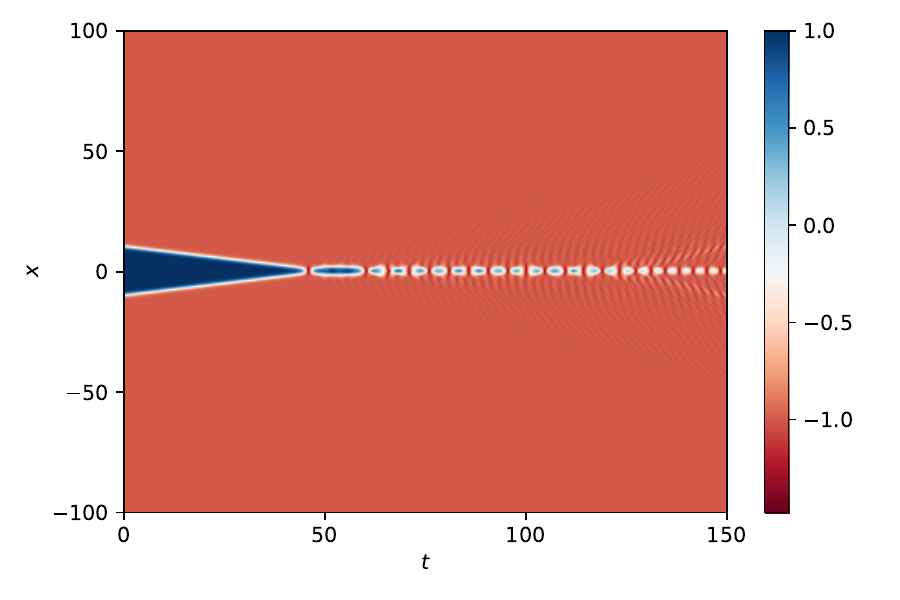}
    \includegraphics[scale=0.28]{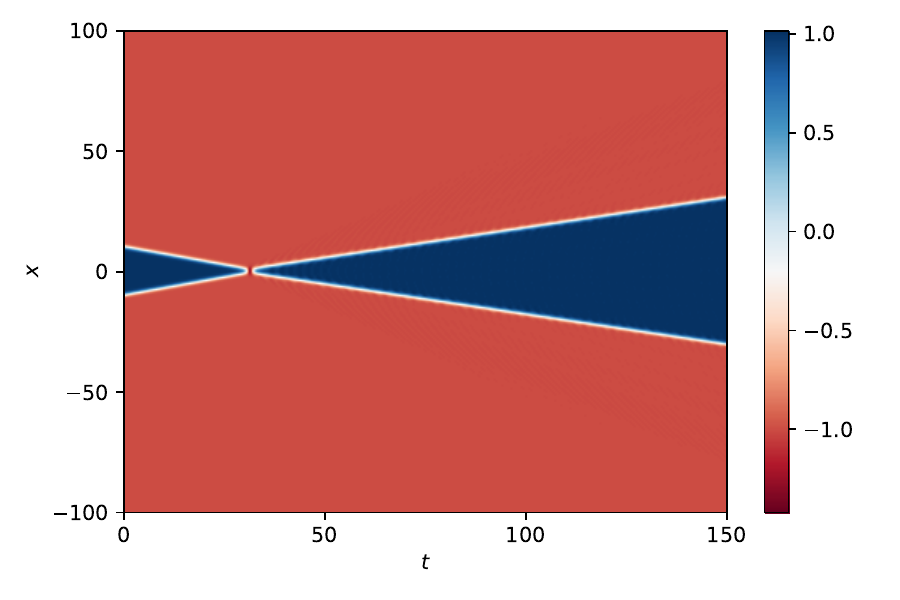}
    \end{center}
    \vspace{-0.5cm}
    \caption{\small{Kink-antikink scattering: field evolution in spacetime for $n=1$ with a) $v=0.20$ and b) $v=0.30$.}
    \label{colB}}
\end{figure}

Fig. \ref{colB}a depicts an approximation of the kink-antikink pair, and after the first collision, an oscillatory mode is created, followed by radiation. In this case, the oscillation amplitude varies between the minima of the potential, and after a long period of evolution, the pair annihilates. This outcome differs when high initial velocities are considered. Fig. \ref{colB}b depicts the case of a single collision and separation of the kink-antikink pair. As this is a non-integrable model, radiation is also seen after the first impact.

\begin{figure}[ht]
    \begin{center}
        \includegraphics[scale=0.4]{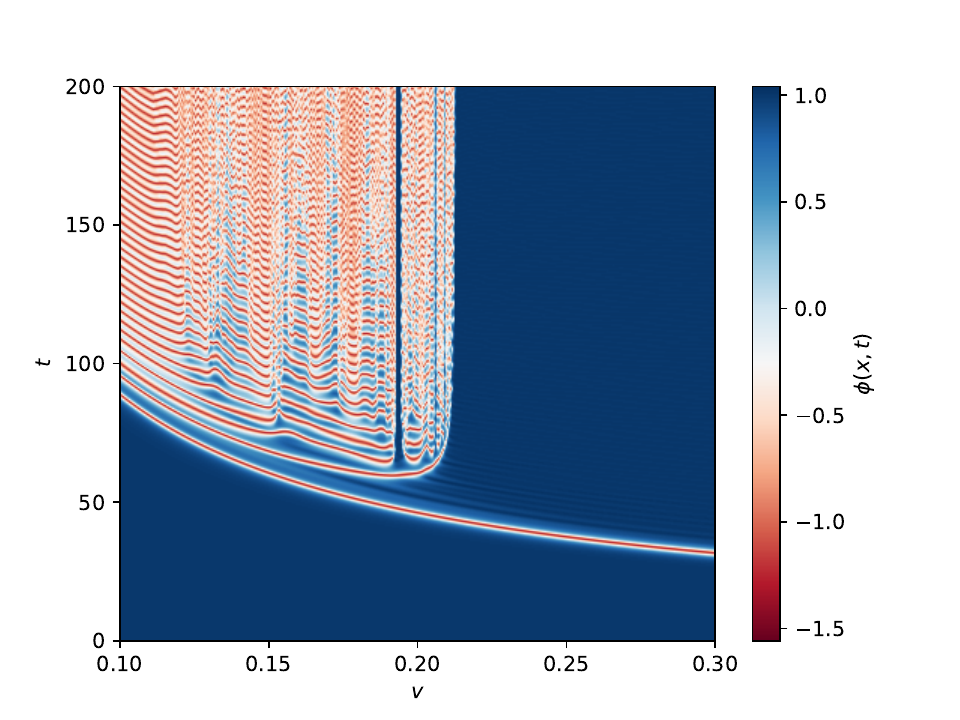}
        \includegraphics[scale=0.4]{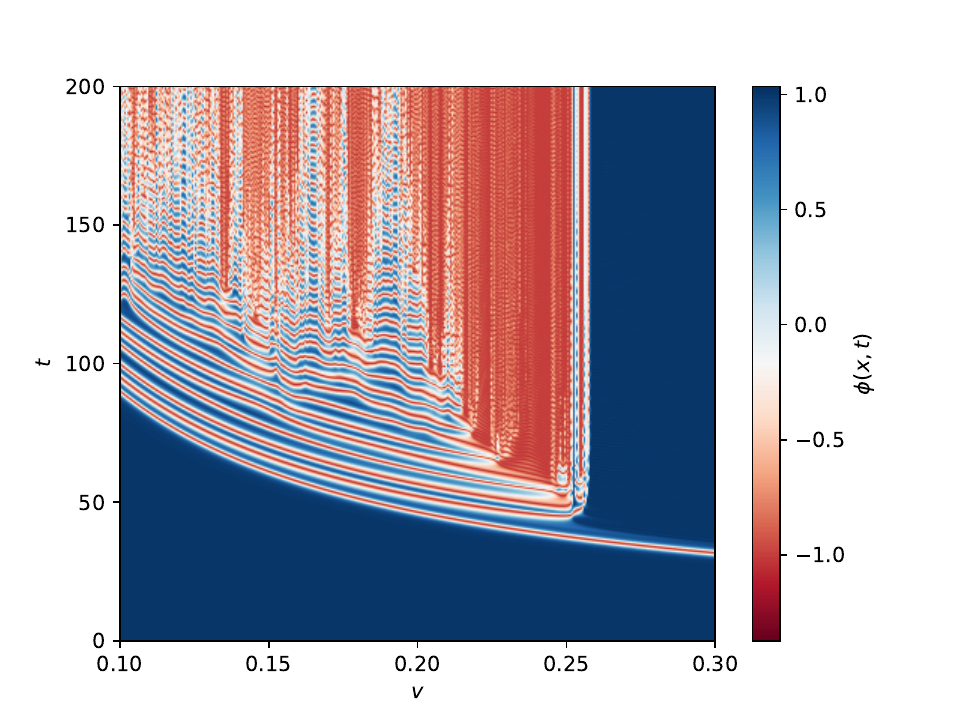}
        \includegraphics[scale=0.4]{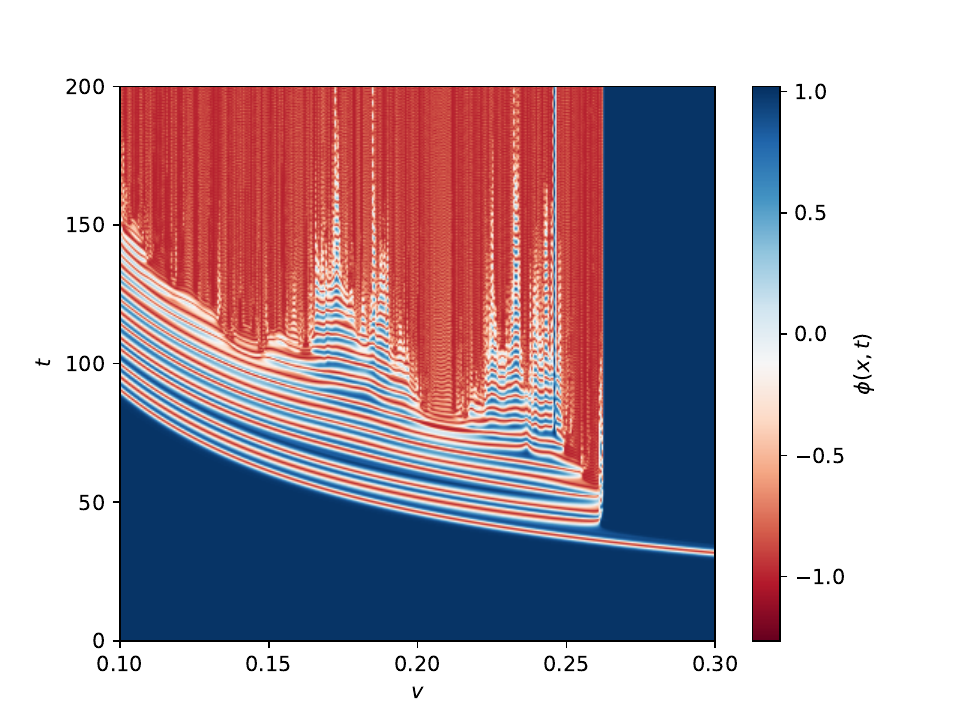}
    \end{center}
    \vspace{-0.5cm}
    \caption{\small{First model - Kink-antikink scattering: evolution of scalar field at the center os mass as a function of time and initial velocity for $n=1$ (top), $n=2$ (middle) and $n=3$ (bottom).}
    \label{colC}}
\end{figure}

We performed several collisions for some values of $n$ and present them in Fig. \ref{colC}. This figure represents the time between bounces as a function of the initial velocity of the kink-antikink pair for $n=1$, $2$ and $3$. It is worth noting that for $n=1$, there are still thin two-bounce windows, which are indicated by thin blue vertical lines. For small velocities, there is only bion-type behavior; however at high velocities, there is only a blue zone, which corresponds to the separation of the pair after a collision. As the parameter $n$ increases, we observe both the suppression of two-bounce windows and a slight increase in critical velocity. The presence of a vibrational mode suggests a resonant window structure, as mentioned in Ref. \cite{Campbell:1983}. However, as the number of states increases, the resonant energy exchange mechanism becomes more complicated, resulting in the suppression of two-bounce behavior \cite{Simas:2016}.

\begin{figure}[ht]
    \begin{center}
        \includegraphics[scale=0.4]{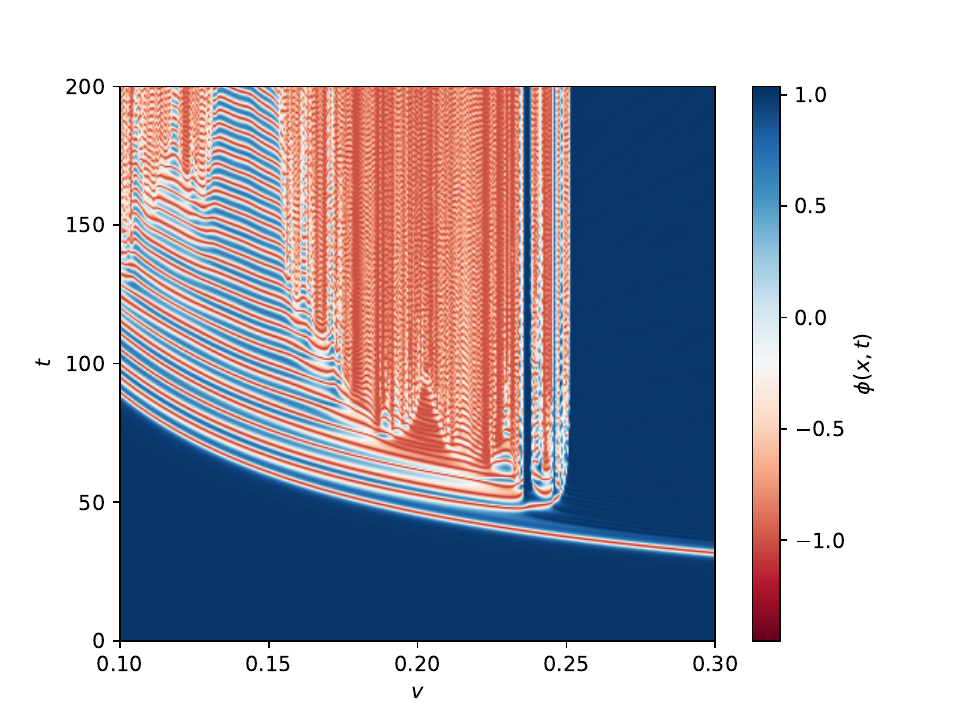}
    \end{center}
    \vspace{-0.5cm}
    \caption{\small{First model - Kink-antikink scattering: evolution of scalar field at the center os mass as a function of time and initial velocity for $n=1.5$.}
    \label{colC1}}
\end{figure}

As we can see, half-integer values of $n$ are also allowed, leading to potentials that are smooth functions of the scalar field. For this reason, we have also examined kink-antikink scattering for such values. It is important to mention that, for $n=0.5$ we get back to the $\phi^4$ model. Additionally, we performed collisions for the case $n=1.5$, as can be seen in Fig. \ref{colC1}. In this case, we see that the structure of resonant windows decreases considerably, showing a wide band with bion-type behavior, followed by a single collision. This result is similar to what occurs for $n=1$, however, for $n=1.5$, there is an increase in the value of the critical velocity. 

{IV.B. {\it Second model. --}
Kink-antikink and antikink-kink scattering processes will be examined in this section. The reason for this is that for a single kink or antikink, linear perturbation analysis produces an asymmetric potential for $n\geq 1$. This behavior is observed in Ref. \cite{Dorey:2011}.

We employed the identical numerical procedure as previously described for model I. For the kink-antikink case, we used the following initial conditions
\begin{eqnarray}
\phi(x,0,x_0,v) \!&=&\! \phi_{K}( x+ x_{0},0,v)  + \phi_{\bar{K}}( x-x_0,0,-v) -1 \text{,}\label{ic3} \nonumber
\\ 
\dot{\phi}(x,0,x_0,v) \!&=&\! \dot{\phi}_{K}( x+ x_{0},0,v ) + \dot{\phi}_{\bar{K}}(x-x_0,0,-v )\text{,} \nonumber
\end{eqnarray}
where $\phi_K(x,t,v) = \phi_{K}\big(\gamma(x-vt) \big)$ means a boost of Lorentz for the static kink solution, with $\gamma=(1-v^2)^{-1/2}$ and $v$ representing the velocity. 

First, we investigated the collision process in the scenario where $n=0$. In particular, there are both translational and vibrational modes in this case. The resonant structure for $n=0$ is displayed in Fig. \ref{colD}, where the horizontal axis represents the initial velocity of the pair and the vertical axis is the collision time. We can see from this figure that the blue lines correspond to the collision of the kink-antikink pair. The yellow region corresponds to the separation of the pair after the interaction. For example, for $v=0.18$, we see two horizontal blue lines and then a wide vertical yellow stripe, which means the separation of the pair after two collisions. In addition, as the velocity $v$ increases, the second blue line diverges, leaving only one collision visible.

\begin{figure}[ht]
    \begin{center}
        \includegraphics[scale=0.4]{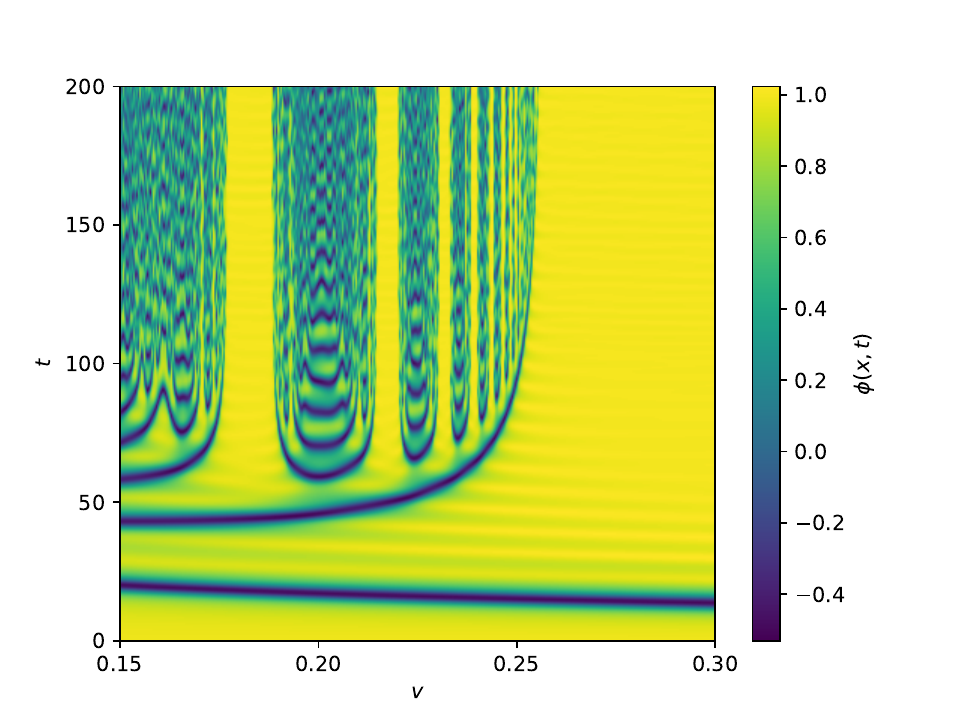}
    \end{center}
    \vspace{-0.5cm}
    \caption{\small{Second model - Kink-antikink scattering: evolution of scalar field at the center of mass as a function of time and initial velocity for $n=0$.}
    \label{colD}}
\end{figure}

The formation of the two-bounce window structure for $n=0$ is thus attributed to the resonant energy exchange mechanism. However, it is worth remembering that increasing the value of $n$ produces a potential with only one zero mode. As a result, the scattering process for $n\geq 1$ exhibits just bion-type behavior and still one-bounce for large values of $v$.

From now on, we present the results of the antikink-kink collision. The initial conditions for this case are as follows
\begin{eqnarray}
\phi(x,0,x_0,v) &=& \phi_{\bar{K}}( x+x_{0},0,v)  + \phi_{K}( x-x_0,0,-v)\text{,}\label{ic3} \nonumber
\\ 
\dot{\phi}(x,0,x_0,v) &=& \dot{\phi}_{\bar{K}}( x+ x_{0},0,v ) + \dot{\phi}_{K}(x-x_0,0,-v )\text{,} \nonumber
\end{eqnarray}
where $\phi_{K}( x+ x_{0},t,v)$ means a boost of Lorentz for the static kink solution, with $\gamma=(1-v^2)^{-1/2}$ and $v$ representing the velocity. 

\begin{figure}[ht]
    \begin{center}
        \includegraphics[scale=0.4]{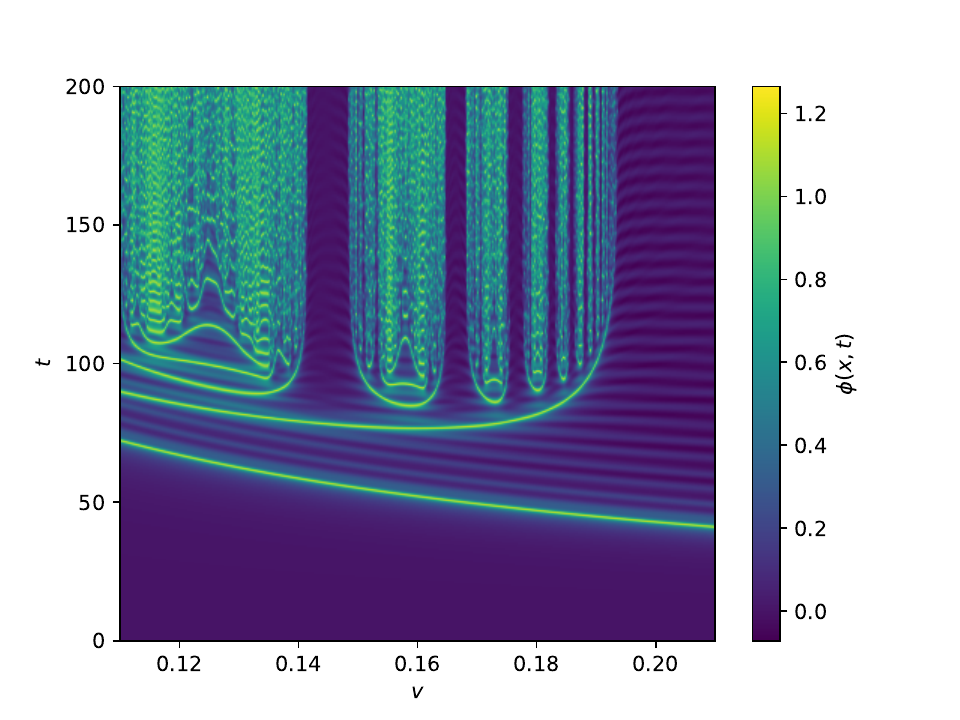}
        \includegraphics[scale=0.4]{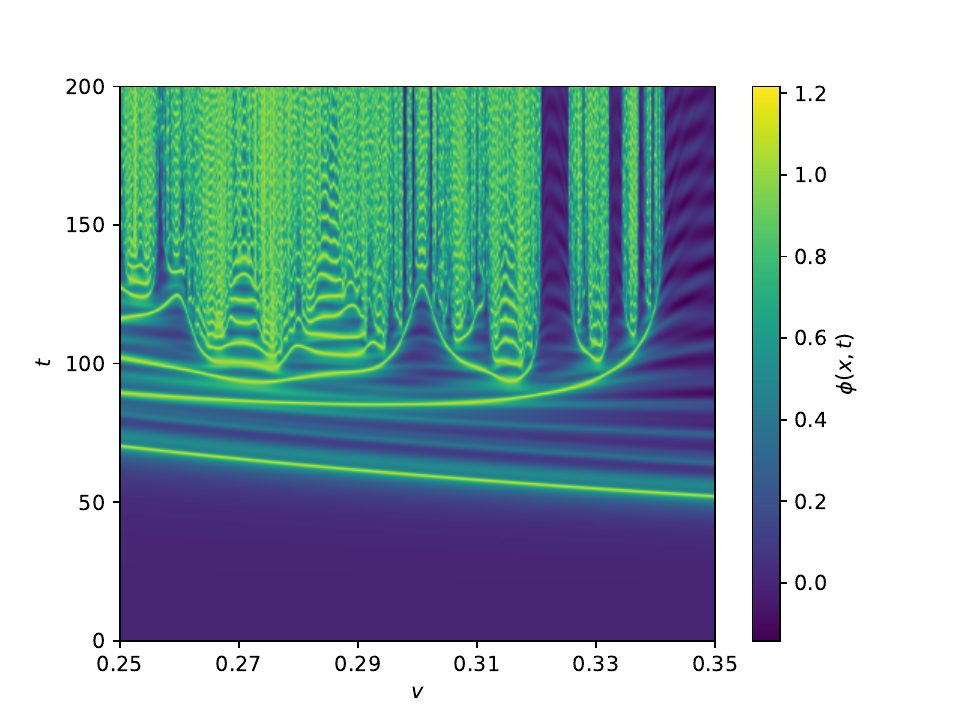}
        \includegraphics[scale=0.4]{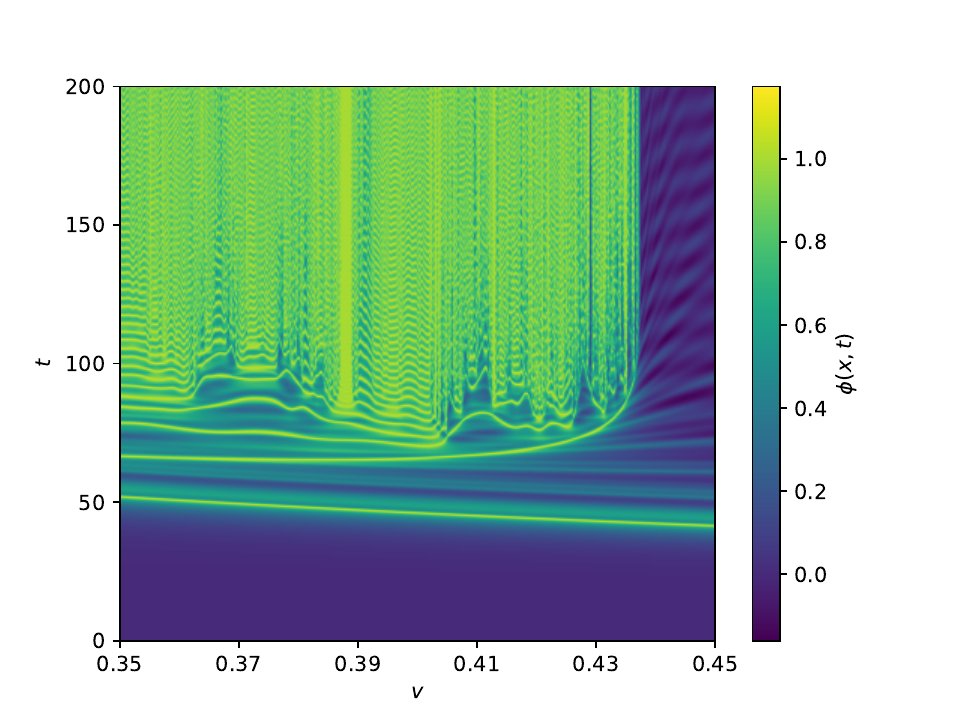}
    \end{center}
    \vspace{-0.5cm}
    \caption{\small{Second model - Antikink-kink scattering: evolution of scalar field at the center of mass as a function of time and initial velocity for $n=1$ (top), $n=2$ (middle) and $n=3$ (bottom).}
    \label{colE}}
\end{figure}

In the current case, resonant window structures are also present, as in the kink-antikink case for $n=0$. On the other hand, two-bounce windows are suppressed when $n$ is increased. Fig. \ref{colE} depicts the interval between collisions as a function of initial velocity for $n=1$, $2$ and $3$. From this figure, we can characterize the green lines as antikink-kink collisions, while the blue area corresponds to the separation of the pair. The two-bounce windows are represented by the vertical blue lines in Fig. \ref{colE} (top) that run between the velocity range $v=0.14$ and $v=0.19$. Note that when $v$ increases, the thickness of these windows decreases and leads to the divergence of the second green line, revealing one-bounce collisions. Furthermore, each window can be identified by the number of oscillations $m$ between two bounces. For example, the first resonant window visible in Fig. \ref{colE} (top) corresponds to the second two-bounce window ($m=2$). This indicates that the first window has been suppressed. The explanation for this result is related to the high number of vibrational states, which hinders the resonant energy exchange process. 

The impact of additional bound modes becomes more evident as $n$ increases, as the two-bounce windows are more suppressed; see Figs. \ref{colE} (middle and bottom). In addition, the area occupied by bion-type behaviors is increasing.

\begin{figure}[ht]
    \begin{center}
    \includegraphics[scale=0.3]{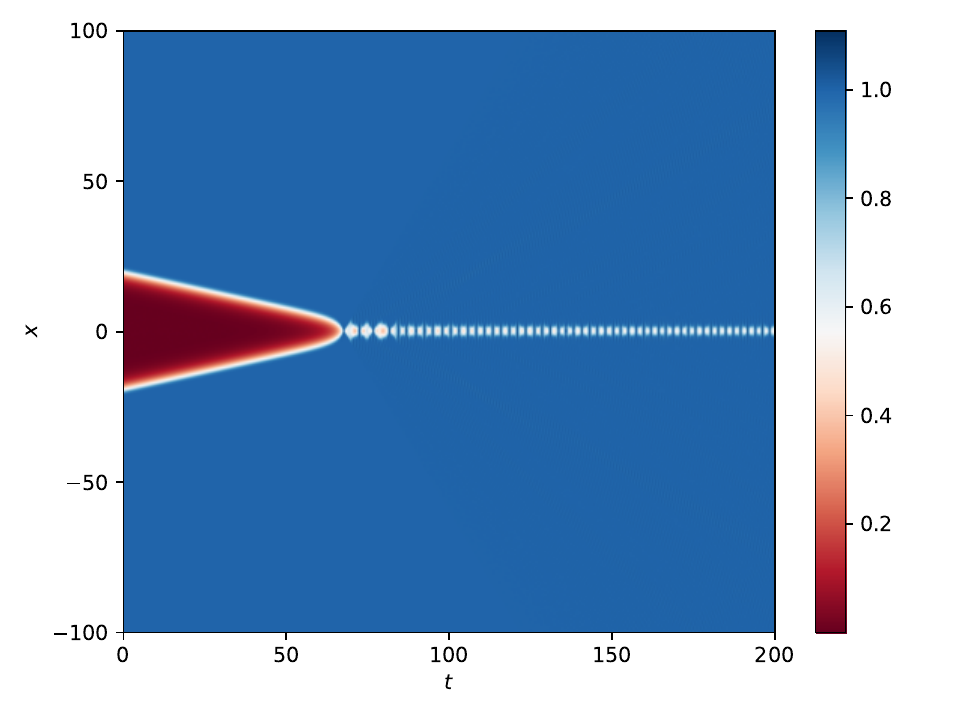}
    \includegraphics[scale=0.3]{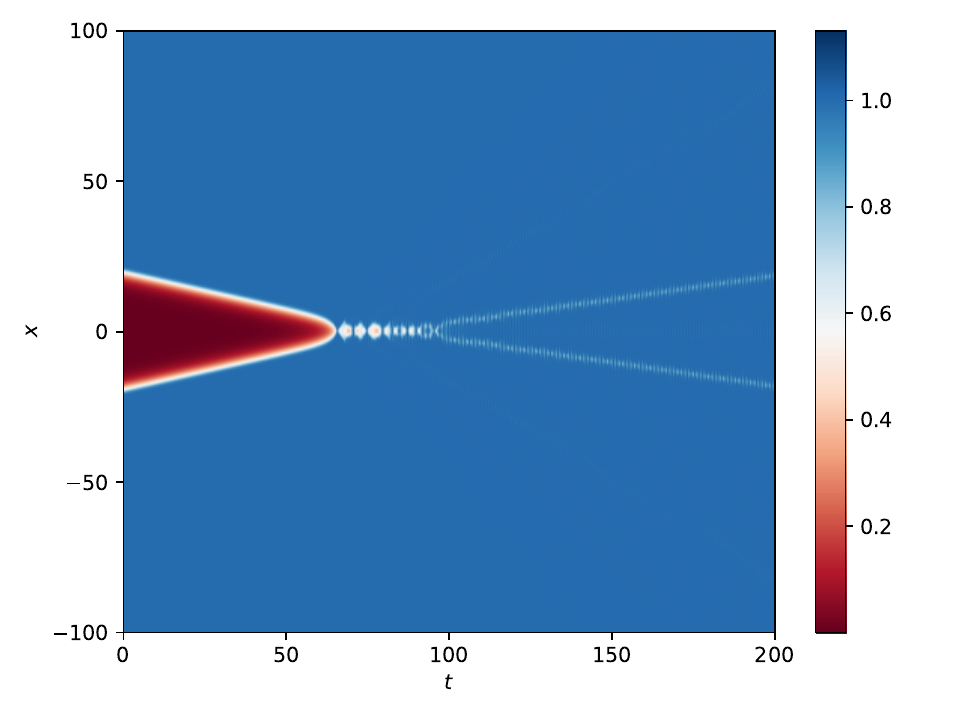}
    \includegraphics[scale=0.3]{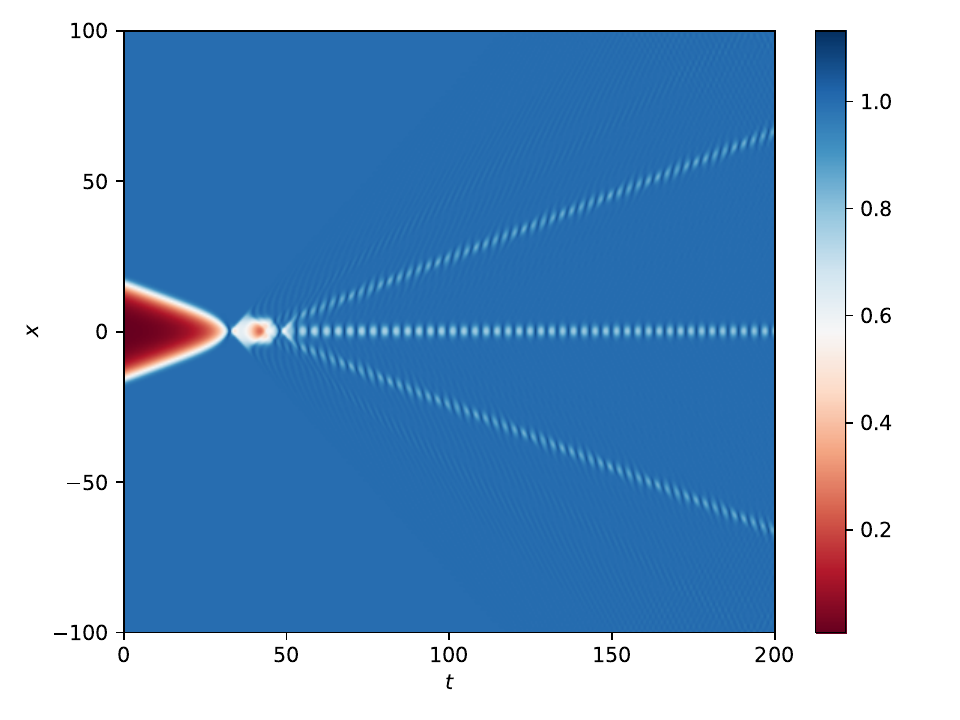}
    \end{center}
    \vspace{-0.5cm}
    \caption{\small{Antikink-kink scattering: field evolution in spacetime for $n=3$ with a) $v=0.1185$, b) $v=0.1225$ and c) $v=0.30$.}
    \label{colF}}
\end{figure}

On the other hand, an interesting aspect of the increase in bound states is the scattering of oscillating pulses after the collision. For large values of $n$, we may observe the emergence of these oscillating solutions, which have a higher harmonicity than bion-type behavior and a longer lifetime. For example, Fig. \ref{colF} depicts the formation of this type of structure. It is important to note that the initial velocity has a significant influence on the result of the antikink-kink collision. In Fig. \ref{colF}a, an oscillatory state is reached after the pair collide. A change in $v$ provides two other types of results, a two-pulse scattering (Fig. \ref{colF}b) and another with three oscillating pulses (Fig. \ref{colF}c). These behaviors are somewhat similar to those found in Refs. \cite{Gani2018_1,Gani2018_2}.

\begin{figure}[ht]
    \begin{center}
        \includegraphics[scale=0.4]{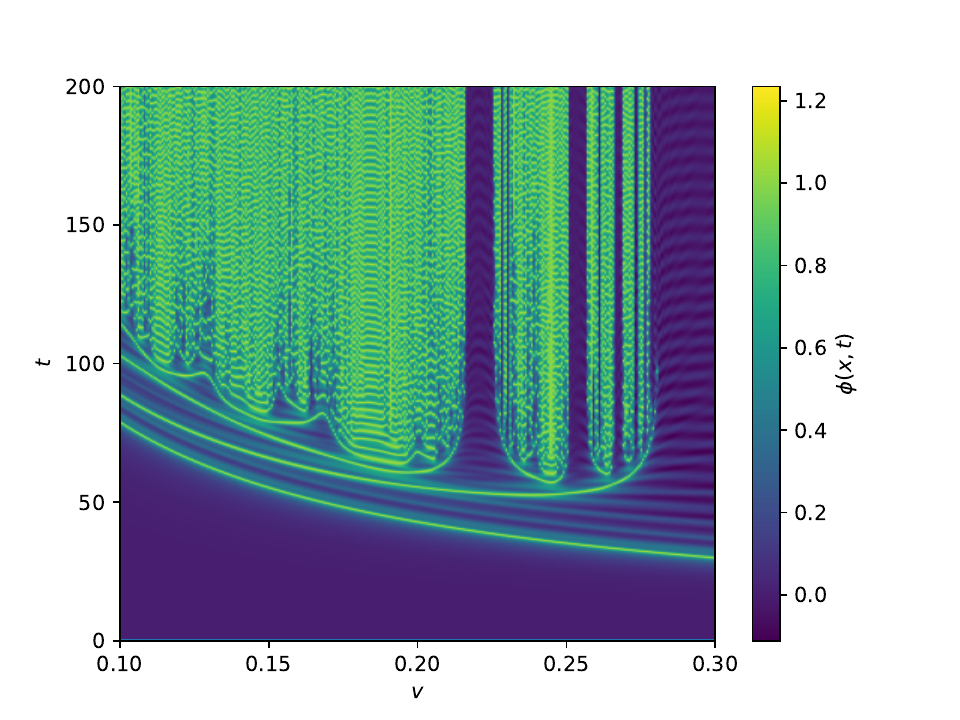}
    \end{center}
    \vspace{-0.5cm}
    \caption{\small{Second model - Antikink-kink scattering: evolution of scalar field at the center of mass as a function of time and initial velocity for $n=1.5$.}
    \label{colG}}
\end{figure}

It is important to note that even for non-integer values of $n$, the potentials are well defined. Based on this, we also investigated the kink scattering process for other values on $n$. For instance, for $n=0.5$, we get a result that is comparable to the $\phi^6$ model for antikink-kink scattering. We also performed several collisions for the case where $n=1.5$ and show them in Fig. \ref{colG}. From this figure, it is also possible to see the presence of resonance windows clustering around a critical velocity, whose value is higher than in the case of $n=1$. Therefore, this result corresponds to a transition between what happens for the values $n=1$ and $n=2$, as we can see in Figs. \ref{colE}a and \ref{colE}b, respectively.

We investigated the scattering process for both models for larger values of $n$, for instance, $n=6,7$, which leads to a compact-like kink solution. The results indicated an increase in the annihilation region of the pair through bion-like behavior and oscillons. The high number of bound states is another element that significantly contributes to this kind of outcome, which makes it more difficult to realize the mechanism of resonant energy exchange between the modes.

{V. \it Ending comments. --}
In this work, we investigated kink solutions in one-dimensional scalar field models, with a particular focus on how different potential configurations influence both the intrinsic properties and scattering processes of these solutions. Using the BPS formalism and linear stability analyses, we explored two models: one with two minima and another with three minima in the potential, both parameterized by an integer parameter $n$.

The results demonstrate that introducing the absolute value of the scalar field in the potential of the models significantly alters both the static behavior of the kink solutions and their dynamic interactions. We observed that as the parameter $n$ increases, the solutions become more compact, each in its own way, with a higher concentration of energy in specific regions. In the first model, characterized by two minima in the potential, we observed that the compactification resulted in a more concentrated energy density around the core of the kink and altered the stability potential, introducing a large number of bound states. The scattering dynamics for this model showed that the resonant structures persisted for lower $n$ values, but were progressively suppressed as $n$ increased.
In the second model, with three potential minima, the inclusion of the absolute value introduced an asymmetric behavior in the investigated quantities, such as the energy density and the stability potential, which became primarily confined to one side of the solution. For lower $n$, the kink-antikink interactions exhibited well-defined resonance windows, while for higher $n$, these windows were replaced by bion-like structures or oscillatory pulses.

In the two models investigated in the present work, we found that the resonant structures as the two-bounce windows tend to disappear as $n$ increases. This phenomenon is attributed to suppression of the resonant energy exchange mechanism caused by the presence of multiple vibrational modes in the stability potential. Furthermore, in collisions involving high values of $n$, oscillatory pulse solutions emerge, showing distinct dynamic behaviors.

These findings demonstrate that modifying the potential through the inclusion of the absolute value of the scalar-field allows for the redesign of kink solutions with new behaviors, opening new possibilities for theoretical and applied studies. In this context, future work could explore extending these analyses to higher dimensions, the inclusion of additional fields to investigate the interplay between different topological structures or even the inclusion of noncanonical kinetic terms, similar to what was done in recent investigations \cite{Marjaneh:2024yec,Lozano-Mayo:2021cfx, Sebastian:2023dds}. We can also think of considering the above results in connection with the presence of stable kink-antikink bound states in a generalized Ginzburg-Landau equation, as reported previously in Ref. \cite{Malo}, in a model that supports self-interactions of the fourth and sixth order. In this context, it may be directly connected with the model engendering three minima, which we have just investigated, so it seems to be of interest to further study the model within the context of the Ginzburg-Landau equation in the case of weak dispersion, to see how the stable kink-antikink solutions behave under collision. Some of these ideas are presently under consideration, and we hope to report on them in the near future.

\begin{acknowledgments}

This work was partially supported by Conselho Nacional de Desenvolvimento Cient\'\i fico e Tecnol\'ogico (CNPq, Grants No. 303469/2019-6 (DB) and No. 402830/2023-7 (DB)), by Coordena{ç}\~ao de Aperfei{ç}oamento de Pessoal de N\'\i vel Superior (CAPES, Finance Code 001), and by Funda{ç}\~ao de Amparo \`a Pesquisa e ao Desenvolvimento Cient\'\i fico e Tecnol\'ogico do Maranh\~ao (FAPEMA, Grant no. 07838/17 (FCS)).

\end{acknowledgments}


\end{document}